# Compact, Reconfigurable Optical Delay Line on a Bent Silica Fiber

MANUEL CRESPO-BALLESTEROS,[1,][*] AND MISHA SUMETSKY[1]

[1]*Aston Institute of Photonic Technologies (AiPT), Aston University, Birmingham, B4 7ET, UK*
*\**m.crespo@aston.ac.uk*

**Abstract:** Tunable optical delay lines that simultaneously offer nanosecond-scale delay, broadband operation, low dispersion, and compact footprint remain challenging to realize with conventional integrated photonic platforms. Here we demonstrate a mechanically reconfigurable slow-light delay line based on a surface nanoscale axial photonics (SNAP) microresonator dynamically induced by controlled bending of a silica optical fiber. A localized nanoscale cutoff-wavelength dip, introduced by $CO_2$-laser annealing, provides a reflective boundary, while fiber bending generates a smooth axial potential whose shape is continuously tunable via loop curvature. By adjusting the bending radius, the induced SNAP microresonator evolves from a nearly linear to an approximately semiparabolic axial profile, enabling controlled transition from dispersive to nearly dispersionless delay. Using a transverse microfiber coupler operated at the impedance-matching condition, we experimentally demonstrate continuous delay tuning from 2 ns to 0.5 ns within a 10 GHz bandwidth in a ~2 mm-long fiber segment, with insertion loss below 6 dB. The results confirm that mechanically induced SNAP microresonators provide a compact, robust, and reconfigurable platform for dispersion-engineered optical delay lines, with direct relevance to photonic beamforming, frequency-comb stabilization, and neuromorphic photonic signal processing.

## 1. Introduction

An optical delay line is a device or system that introduces a controlled time delay to a light signal as it propagates from one point to another. Fundamentally, it operates by increasing the optical path length or reducing the group velocity of light in a given medium, thereby slowing down the signal relative to free-space propagation. Delay lines are essential components across various advanced technologies where precise timing control and synchronization of optical signals are required. In radar and LiDAR beam steering, precise control of optical delay lines enables dynamic adjustment of signal phases for accurate beam directionality and adaptive scanning [1–6]. Optical frequency comb (OFC) stabilization benefits significantly from optical delay lines by providing precise control of pulse timing, leading to highly stable frequency combs essential for spectroscopy and high-precision metrology [7–9]. Tunable optical delay lines play an important role in coherent optical communication systems, where they help align signals and correct for dispersion effects that can distort data at high speeds [10,11]. In quantum photonics, delay lines are used to synchronize single photons generated at random times, making it possible to build more reliable and scalable quantum networks [12–14]. In neuromorphic photonic systems, especially reservoir computing, delay lines are used to enhance computational capabilities by creating memory elements within photonic circuits, significantly reducing computational complexity and energy consumption compared to electronic alternatives [15–18].

Optical delay lines can be implemented through several physical mechanisms. The most direct method involves increasing the physical length that light travels, such as in optical fiber coils or waveguide spirals [19–21]. This geometry-based method, however, becomes impractical when nanosecond-scale delays are required in compact or on-chip systems. More advanced methods exploit the phenomenon of slow light, where the group velocity of a light pulse is significantly reduced due to resonant or structural effects within the material or device [22–24].

Slow light allows for large temporal delays over short distances, enabling compact and efficient photonic delay lines. Slow light platforms include microring resonators, coupled-resonator optical waveguides (CROWs) [25,26], and side-coupled integrated spaced sequences of resonators (SCISSORs) [27,28]. These technologies offer several nanosecond delays in relatively compact footprints, but at the cost of limited bandwidth, dispersion, loss, and restricted real-time tunability. They also demand sub-nanometre dimensional tolerances: an error of ~1 nm in waveguide width or resonator gap can shift a resonance by 100 pm or even 1 nm [29,30], which detunes cascaded resonators, degrades the engineered group-index profile, and decreases the slow-light effect on which the delay line relies. Achieving sub-angstrom level precision uniformly across a wafer remains beyond standard lithographic processes, adding complexity and cost.

Surface nanoscale axial photonics (SNAP) microresonators provide an attractive alternative to current delay lines technologies. SNAP microresonators are high-Q whispering gallery mode (WGM) microresonators [31–33]. They can significantly reduce footprint, simplify fabrication, and enhance robustness and tunability compared to conventional integrated photonic platforms. Their high Q-factor and nanoscale precision ensure ultra-low propagation losses, enabling extremely slow propagation of light ("slow light") along the fiber axis. Slow light refers to a substantial reduction in group velocity, allowing significant temporal delay within a compact footprint, an advantage for delay line applications where timing precision and device size are critical. In SNAP structures, this effect arises from the axial propagation of WGMs, which circulate azimuthally around the fiber while gradually propagating along the fiber axis. The resulting axial group velocity can be orders of magnitude lower than the speed of light in the fiber, enabling nanosecond-scale delays over millimeter-length segments [34–36].

As shown in Ref. [34], a semiparabolic SNAP microresonator yields a set of equidistant eigenfrequencies and hence supports dispersionless delay, making it an ideal geometry for slow light delay lines. In our recent work [37], we demonstrated that a parabolic SNAP microresonator can be precisely induced and tuned by mechanically bending a silica optical fiber with sub-micron control. By adjusting the bending radius, the effective axial potential, and hence the free spectral range (FSR), can be continuously tuned over a significant range with sub-picometer precision. This approach enables the creation of dispersion-engineered SNAP devices without permanent structural modification, offering a compact, repeatable, and reconfigurable platform. Building on this bending technique, in this paper we present a tunable SNAP microresonator delay line of the type proposed in Ref. [34], but with the added capability of real-time tunability.

## 2. Theoretical background

In SNAP microresonators, WGMs are localized near the surface of a silica optical fiber, where they are confined transversely by total internal reflection. These modes exhibit strong radial and azimuthal confinement and propagate slowly along the fiber axis in response to nanoscale variations in the fiber radius or refractive index. The axial propagation of WGMs can be effectively modeled using the concept of *effective radius variation* (ERV) $\Delta r(z)$, which captures the impact of physical radius changes and refractive index perturbations on the local cutoff wavelength. This ERV is related to the cutoff wavelength variation $\Delta \lambda(z)$ by the scaling relation $\Delta r(z) = r_0 \Delta \lambda(z)/\lambda_0$, where $r_0$ is the unperturbed fiber radius and $\lambda_0$ is the resonance wavelength of the unperturbed WGM. SNAP microresonators are fabricated by precise nanoscale manipulation of the ERV $\Delta r(z)$ along the axial direction $z$ of a silica optical fiber [37–43].

Like in other WGM microresonators, light propagation in SNAP microresonators is governed by the scalar Helmholtz equation, where the refractive index varies slowly along the fiber axis due to nanoscale geometric or refractive index perturbations already mentioned. In the case of a straight or weakly bent fiber, the structure exhibits approximate cylindrical symmetry, allowing the use of cylindrical coordinates $(\rho, \phi, z)$. Under these conditions, the light field can be factorized as $\Psi(\rho, \phi, z) = \Phi(\rho)e^{im\phi}\psi(z)$, where $\Phi(\rho)$ describes the radial field distribution, $e^{im\phi}$ captures the azimuthal dependence with *quantum* number $m$, and $\psi(z)$ is a slowly varying envelope along the fiber axis. This separation is justified by the adiabatic approximation, which assumes that axial variations in the fiber geometry occur over a length scale much larger than the transverse mode size and the optical wavelength, allowing the transverse mode to adapt quasi-statically to the local geometry.

However, in the presence of strong bending, such as in the SNAP microresonator of Ref. [37] and in the case of the device proposed in this paper, global cylindrical symmetry is no longer applicable. In this case, the axial coordinate $z$ must be replaced by the arclength coordinate $s$ that follows the bent fiber axis (Fig. 1). The transverse structure is then defined in the local normal plane perpendicular to the tangent vector $\vec{t}$ of the fiber at each point along $s$. In our configuration, the fiber loop radius $R_f(s)$ is few mm, while the transverse mode size is approximately $w_t \sim 1.5\,\mu m$, yielding a ratio $R_f(s)/w_t \sim 1000$. This large separation of scales ensures that the transverse mode evolves adiabatically along the arclength coordinate $s$ and a local separation of variables remains valid. For wavelengths $\lambda$ near a resonance cutoff wavelength $\lambda_0$, the axial field envelope $\psi(s)$ satisfies the one-dimensional Schrödinger-type equation [31,44]:

$$\frac{d^2\psi(s)}{ds^2} + [E(\lambda) - V(s)]\psi(s) = 0, \qquad (1)$$

For clarity, we retain the term *axial* even though $s$ follows the curved fiber axis. The axial confinement of the WGM is due to the spatial variation of $V(s)$, which defines an effective potential well in Eq. 1. This potential arises from nanoscale variations in the effective radius, as well as curvature- and stress-induced shifts in the local cutoff wavelength. As in the straight case, the potential can be expressed as $V(s) = -\kappa^2 \Delta r(s)/r_0$, with ERV $\Delta r(s) = r(s) - r_0$ and $\kappa = 2^{3/2}\pi n_0/\lambda_0$. The effective energy is given by $E(\lambda) = -\kappa^2 \Delta\lambda/\lambda_0$ with $\Delta\lambda = \lambda - \lambda_0 - i\gamma$, where $\gamma$ is the attenuation parameter in silica (typically <0.1 pm).

The local axial propagation constant $\beta(\lambda, s)$ governing phase accumulation is defined via the WKB approximation:

$$\beta(\lambda, s) = \sqrt{E(\lambda) - V(s)} \qquad (2)$$

In SNAP devices, both $E(\lambda)$ and $V(s)$ are small quantities near resonance, and as a result, the axial propagation constant $\beta$ is much smaller than in typical fiber modes, often on the order of $10^3\,m^{-1}$ or less (in comparison with $10^6\,m^{-1}$ in conventional fibers). This small value of $\beta$ underlies the ultraslow axial group velocity observed in SNAP microresonators.

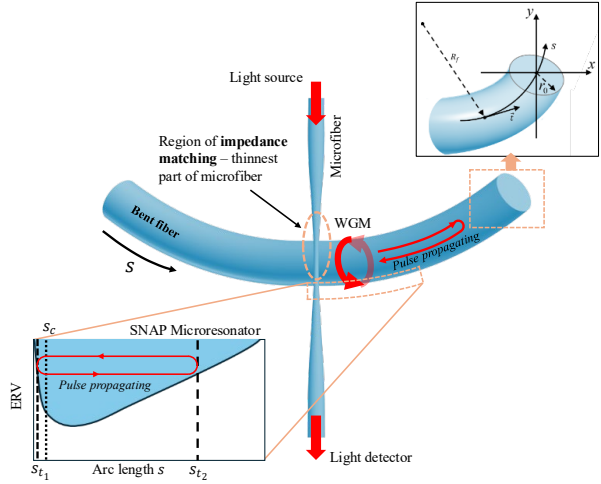

Fig. 1. Illustration of the proposed device. A dip in the ERV (or CWV) is introduced in the bent fiber using a $CO_2$ laser. The bending process induces a SNAP microresonator with an end face (inset) that can be used as a delay line.

Light is coupled into and out of the SNAP microresonator at position $s_c$ using a biconical microfiber taper with a waist diameter of $\sim 1\ \mu m$ (Fig. 1). In the next section, we will describe how we engineered the ERV to obtain a SNAP microresonator with an end face similar to the semiparabolic microresonator of Ref. [34] and sketched in the inset of Fig. 1. The axial turning points $s_{t1}, s_{t2}$ are located at positions where $\beta(\lambda, s) = 0$, and the supported axial modes correspond to bound states of the potential $V(s)$. In this regime, the axial component of the WGM propagates with an effective group velocity determined by the spectral derivative of the local propagation constant. In SNAP structures, this axial group velocity can be reduced to values as low as $0.001c$, enabling nanosecond-scale delays over fiber segments only a few millimeters long. The transmission amplitude $S(\lambda, s_c)$ through the microfiber and the corresponding group delay $\tau(\lambda, s_c)$ associated are:

$$S(\lambda, s_c) = S_0 - \frac{i|C|^2 G(\lambda, s_c, s_c)}{1 + DG(\lambda, s_c, s_c)} \quad (3a)$$

$$\tau(\lambda, s_c) = \frac{\lambda^2}{2\pi c} Im\left\{\frac{d}{d\lambda} \ln S(\lambda, s_c)\right\} \quad (3b)$$

where $G(\lambda, s_c, s_c)$ is the Green's function of the uncoupled axial Schrödinger problem, $S_0$ is the out-of-resonance transmission amplitude, and $C$ and $D$ are complex coupling parameters that characterize the strength and phase of microfiber–microresonator interaction [31,38].

Axial confinement and reflection of light within the resonator lead to oscillations in both $S(\lambda, s_c)$ and $\tau(\lambda, s_c)$. These arise from partial reflection of the WGM at the coupling point, causing the pulse to bounce multiple times between axial turning points before being fully extracted. In the spectral domain, this results in periodic modulation due to interference between multiply reflected components. The average group delay over one oscillation period corresponds to the classical round-trip time between turning points and is given by [34]

$$\langle \tau(\lambda) \rangle = \frac{\lambda_0^2}{\pi c} \int_{s_{t1}}^{s_{t2}} ds \frac{\partial \beta(\lambda, s)}{\partial \lambda} \quad (4)$$

These oscillations are suppressed when the coupling satisfies the local impedance matching conditions derived in Ref. [34]. Under these conditions, a light pulse launched into the resonator from the microfiber at the impedance matching position is fully transmitted and, after a single round-trip, returns entirely into the microfiber output, resulting in smooth, ripple-free group delay spectra.

To ensure distortion-free pulse propagation through the resonator, not only must the group delay be well defined, but the group delay dispersion (GDD) must also be minimized. This requires the resonator to support a set of equally spaced axial eigenmodes, since a uniform spectral spacing leads to a linear frequency-dependent phase response, preserving the shape of the pulse in time. In contrast, any deviation from uniform mode spacing results in nonlinear phase accumulation, leading to temporal broadening and pulse distortion. As shown in Ref. [37], a parabolic (or semiparabolic) axial potential yields a set of equidistant eigenfrequencies and hence supports dispersionless delay, making it an ideal geometry for slow light delay lines.

## 3. Device Fabrication and Characterization

The device was fabricated using an uncoated single-mode silica optical fiber with a cladding radius of $r_0 = 19\ \mu m$. The protective polymer coating was stripped in a hot sulfuric acid bath, and the fiber was thoroughly cleaned using isopropanol and acetone to expose a pristine silica surface suitable for post-processing. The axial uniformity of the fiber was first characterized over a 10 $mm$ segment using an Optical Vector Analyzer (OVA, Luna OVA 5000), which measured the original CWV. As shown in Fig. 2, the CWV profile presented a small axial nonuniformity of approximately $0.027\ nm/mm$, corresponding to an ERV slope of $0.33\ nm/mm$. On this fiber, a steep CWV dip was inscribed at the center of the fiber segment using focused $CO_2$ laser annealing (Fig. 2 inset). The purpose of this localized dip in the CWV is to be a reflective boundary, analogous to the end face in the SNAP microresonators of Refs. [34–36]. During annealing, the fiber was slightly stretched with a motorized linear stage to induce local stress relaxation and nanoscale elongation, resulting in a CWV dip with a depth of approximately $4\ nm$ (equivalent to a $49\ nm$ in ERV).

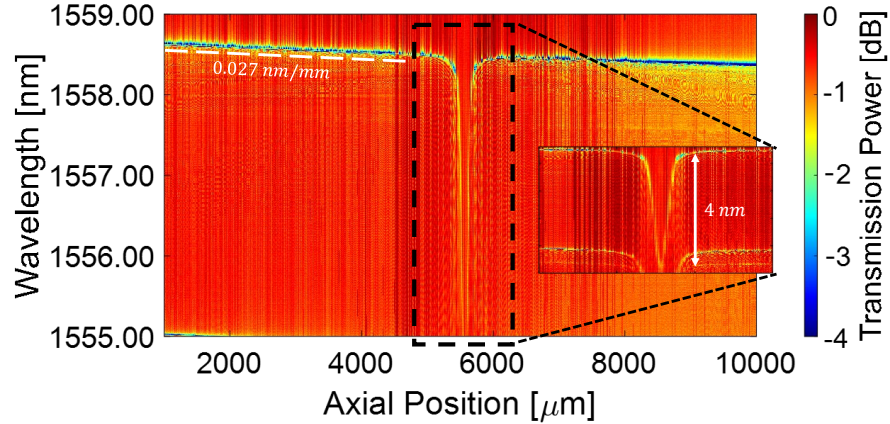

Fig. 2. Spectrogram of the unbent fiber with the introduced CWV dip. Inset: rescaled dip region.

Unlike conventional SNAP microresonators fabricated entirely through permanent CWV shaping, the SNAP microresonator in this work was formed dynamically by bending the fiber into a loop. The fiber was inserted through a narrow plastic capillary tube, with the $CO_2$ inscribed CWV dip placed precisely at the apex of the loop (Fig. 3). The curvature introduced by bending modifies both the fiber geometry and the stress profile: the outer arc of the fiber elongates, while the inner arc compresses. These effects lead to a change of the effective refractive index and radius via geometric deformation and the elasto-optic effect, respectively. The result is a smooth SNAP microresonator centered at the dip location.

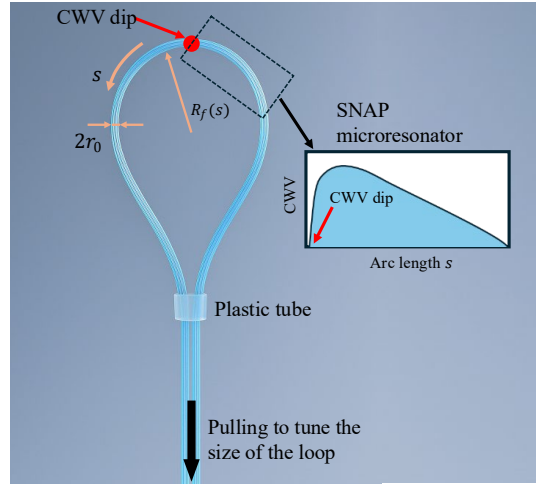

Fig. 3. SNAP microresonator delay line. The CWV dip is placed at the center of the loop. A linear stage is used to pull the fiber and change the size of the loop which in turn changes the shape of the SNAP microresonator and tune the delay.

In this configuration, both ends of the fiber extended beyond the tube and were fixed to a single-axis linear translation stage. By pulling the fiber ends with this stage, the size of the loop, and thus its curvature, was dynamically adjusted. The degree of bending governs the depth and shape of the induced SNAP microresonator, so tunability is achieved by controlling the position of the translation stage and thereby the curvature of the loop. The microfiber taper used to couple light in and out of the microresonator was placed near the CWV dip. The position of the microfiber was finely adjusted to satisfy the impedance matching condition, minimizing spectral oscillations and ensuring efficient light exchange with the axial WGMs.

To characterize the shape of the SNAP microresonator induced by controlled bending of the optical fiber, we measured spectrograms representing the resonance wavelength of the microresonator as a function of the arc length position $s$. To enable precise measurement of these spectrograms, we modeled the geometry of the bent fiber using the elastica theory for inextensible rods in static equilibrium (Fig. 4a) [45]. Specifically, we used the inflectional solution to the elastica problem [46]. The elastica problem reduces to the nonlinear pendulum equation:

$$\frac{d^2\theta(s)}{ds^2} + \xi \sin\theta(s) = 0 \tag{5}$$

where $\theta(s)$ is the angle between the tangent to the fiber at arc length $s$ and the horizontal axis (Fig. 4b). Here, $\xi = F/YI$ is a constant proportional to the ratio of the applied horizontal force $F$ to the bending stiffness $YI$ of the fiber, where $Y$ is the Young's modulus and $I$ is the area moment of inertia of the cross section of the fiber. This equation is mathematically equivalent to the motion of a simple pendulum and captures the balance between internal bending moments and external constraints in a planar, inextensible, and purely elastic rod. By solving this equation, one obtains $\theta(s)$ in terms of Jacobi elliptic functions. The cartesian coordinates of the fiber centerline for the inflectional solution are:

$$x(s) = \zeta \cdot (s - 2E[am(s,m), m]) \tag{6a}$$
$$y(s) = \zeta \cdot \left(-2\sqrt{m} \cdot cn(s,m)\right) \tag{6b}$$

where $E(\phi, m) = \int_0^\phi d\theta \sqrt{1 - m \cdot sin^2\theta}$ is the incomplete elliptic integral of the second kind, $am(s, m)$ is the Jacobi amplitude function, and $cn(s, m)$ is the Jacobi elliptic cosine function. The parameter $m \in (0,1)$ is the elliptic modulus, which parametrizes the shape and curvature of the elastica. $\zeta$ is a scaling factor to adjust the size of the modelled curve to the actual one. These expressions describe a periodic, inflectional elastica curve with points of zero curvature, ideal for modeling the symmetric bent fiber loop used in the experiments (Fig. 4c). The parameter $m$ controls the number and sharpness of inflection points per period, and was chosen in our case to match the observed geometry (Fig. 4b and Fig. 4c).

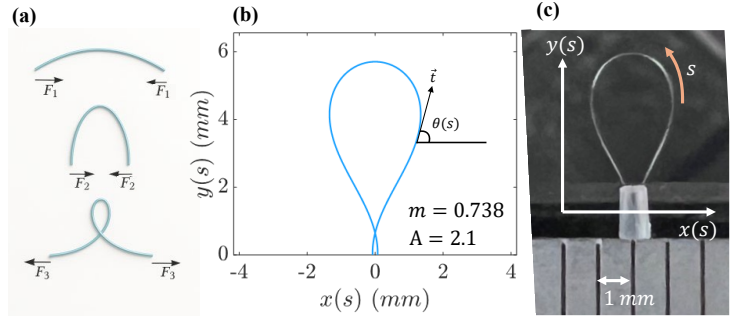

Fig. 4. (a) A slender fiber adopts different shapes depending on the pair of forces applied. These shapes are predicted by the elastica theory. (b) Mathematical model of the bent fiber. This corresponds to the largest loop size $R = 2.6$ mm. The parameters $m$ and $A$ have been adjusted to match the dimensions of the loop used in experiments (c).

The theoretical fiber profile derived from the inflectional elastica solution was implemented in MATLAB to program the scanning trajectory of the microfiber along the bent fiber surface. At each calculated contact point, we acquired the transmission spectrum with an OVA. This approach ensured accurate positioning and consistent coupling conditions throughout the scan, enabling reliable reconstruction of the spectrograms that correlate axial position with resonance wavelength.

The two-dimensional surface plots in Fig. 5 represent the spectrograms obtained for two different loop radii: $R = 2.6\ mm$ and $R = 1.8\ mm$. This provides a direct visualization of the CWV, which determines the effective potential $V(s)$ for WGMs. For the larger loop radius ($R = 2.6\ mm$, Fig. 5a), the spectrogram reveals that a SNAP microresonator is induced by fiber bending. The induced CWV (and, equivalently, the ERV) exhibits a nearly linear profile over the $\sim 2\ mm$ scanned region. The right-hand side of the resonator shows a well-defined end face due to the steep CWV dip initially introduced with the $CO_2$ laser. This linear axial potential results in a dispersive delay line, as will be discussed in the next section.

In contrast, the spectrogram corresponding to the tighter loop ($R = 1.8\ mm$, Fig. 5b) shows that the CWV adopts a more parabolic profile. As will be demonstrated in the following analysis of group delay spectra, this configuration supports nearly dispersionless pulse propagation.

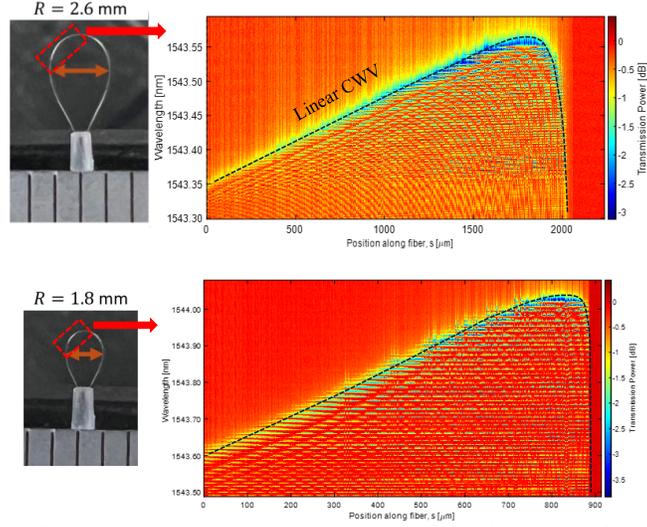

Fig. 5. Spectrograms of the bent fiber. (a) SNAP microresonator corresponding to a bent fiber loop of radius $R = 2.6\ mm$. The induced microresonator exhibits a linear CWV. (b) SNAP microresonator formed by a bent fiber loop of radius $R = 1.8\ mm$. The CWV shows a curved profile that can be approximated by a parabolic curve.

## 4. Experimental Results

We experimentally investigated the tunable delay performance of this SNAP-based device by varying the radius of the bent fiber loop from $R = 2.6\ mm$ to $R = 1.8\ mm$. Fig. 6 shows the measured group delay (top row) and transmission amplitude (bottom row) spectra for different loop radii. The black traces represent the raw experimental data. The red lines correspond to ripple-averaged group delay spectra computed for a 0.1 ns input pulse, and the blue lines correspond to a 0.5 ns pulse. These curves illustrate how the group delay would be experienced by pulses of finite duration, smoothing out rapid spectral oscillations and providing a practical estimation of device performance for different signal bandwidths. The vertical green line in each panel indicates the wavelength $\lambda_0$ of the pulse. As the delay becomes more dispersionless, this wavelength corresponds to minimal spectral ripple due to impedance matching.

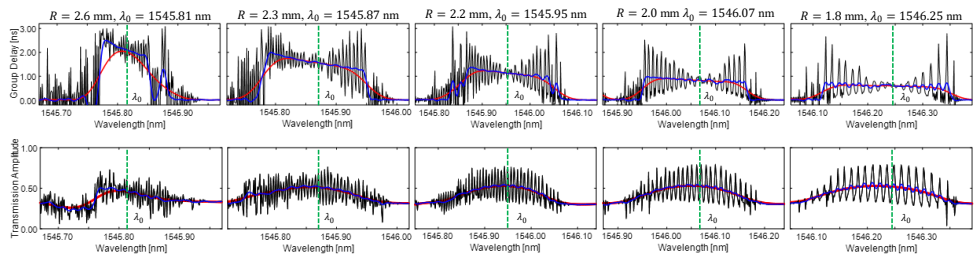

Fig. 6. Group delay (top row) and transmission amplitude (bottom row) spectra measured by the OVA for different loop sizes.

To find the position of the impedance matching condition, we performed high-resolution scans around the CWV dip, moving the microfiber in steps of $0.1\ \mu m$. To fulfill this condition, we increased the coupling losses by shifting the microfiber towards its thinnest region. This increase in coupling losses is necessary to suppress the oscillations in the group delay spectrum and ensure that the input light is fully extracted back into the microfiber after a single roundtrip. According to the theoretical framework based on the renormalized Green's function of the SNAP microresonator [34], the impedance matching condition requires that the imaginary part of the microfiber–microresonator coupling parameter, $Im\{D\}$, equals the local propagation constant at the coupling point, $\beta(\lambda, s)$, and that the coupling strength satisfies $|C|^2 = 2S_0 \cdot Im\{D\}$. These conditions can only be satisfied if the coupling is sufficiently strong, which, in practice, requires enhancing the scattering loss at the coupling point by moving the microfiber to its thinnest part. Once the impedance matching condition was identified, we investigated the tunability of the delay line by fixing the microfiber at this position, in contact with the bent fiber, and reducing the fiber loop radius $R$.

At the largest loop radius ($R = 2.6\,mm$), the device provides a group delay of up to 2 ns, with a measured dispersion of approximately 7.5 ns/nm across a 10 GHz bandwidth. This significant wavelength dependence arises from the axial profile of the microresonator (Fig. 5a), which deviates from the ideal semiparabolic form required for dispersion-free operation. As the loop radius is reduced, the wavelength dependence of the delay becomes progressively flatter. At the smallest radius ($R = 1.8\,mm$), the device achieves a nearly constant group delay of approximately $0.5\,ns$ over a 10 GHz bandwidth. This flattening is consistent with the formation of an approximately semiparabolic CWV (Fig. 5b), which is known to suppress dispersion in SNAP microresonators by maintaining a nearly uniform axial group velocity across the bandwidth.

The transmission amplitude measurements confirm stable coupling and low loss throughout the tuning range. The insertion loss remains below 6 dB, and the resonance bandwidth stays nearly constant at 10 GHz, indicating that the SNAP delay line maintains broadband performance across its tunable range. A small global wavelength shift (~0.4 nm) is observed with decreasing loop radius, corresponding to the effective index change due to bending-induced strain.

## 5. Discussion and Conclusion

The demonstrated tunable delay line based on bent SNAP microresonators provides nanosecond-order delay tunability within a remarkably compact 2 mm fiber segment, achieving a group delay range from 2 ns to 0.5 ns and an operation bandwidth up to 10 GHz. This level of performance, compares favorably with current state-of-the-art integrated delay lines (see Table I in Ref. [25] and references therein).

A key feature of the proposed device is its continuous tunability by mechanical adjustment of fiber loop curvature, which modifies the SNAP microresonator shape and hence its FSR. As the loop radius $R$ decreases, the initially dispersive delay line evolves into a nearly dispersionless configuration, consistent with the formation of a semiparabolic CWV profile [37]. This illustrates how the interplay between CWV design and geometric strain can be used to control the spectral properties of SNAP microresonators in real-time.

The insertion loss observed across the tuning range of the delay line is approximately 6 dB, which, while sufficient for proof-of-concept demonstration, remains relatively high compared to the lower losses demonstrated in previous work [34–36]. These losses are attributed primarily to environmental factors and contamination of the fiber surface, which can degrade the coupling efficiency and increase scattering. Nonetheless, the effective coupling was achieved by careful positioning of the microfiber, as indicated by the suppression of spectral ripples in the group delay response (Fig. 6). The observed spectral shift (~0.4 nm) induced by bending is monotonic and reversible, and could be compensated through uniform thermal tuning, suggesting that integration with temperature stabilization or feedback control may further improve device performance and stability.

Compared to lithographically fabricated photonic delay lines such as CROWs, SCISSORs, or microring arrays, the present approach avoids the complexity, rigidity and lack of precision of planar fabrication, offering instead a flexible and reconfigurable platform. Furthermore, it is possible to combine the laser annealing [42] or the femtosecond inscription fabrication methods [35] with the bend-induced microresonator technique [37,41] for a novel hybrid method of SNAP device engineering. The combination of these different techniques can provide a way to fabricate a large range of delay lines with different delays, dispersion, and wavelength operation.

Future research directions include refining the CWV profiles to achieve ideal parabolic axial potentials that support extended dispersionless operation. This could enhance the performance of SNAP-based delay lines, particularly for high-fidelity signal processing. Another line of research involves integrating actuators or electro-optic elements to enable automated and rapid tuning of the device, thereby expanding its applicability in dynamic photonic systems. Additionally, cascading multiple SNAP microresonators could facilitate longer delay lines or multiplexed architectures, supporting more complex signal routing or temporal processing tasks. Finally, combining SNAP structures with fiber Bragg gratings or planar photonic circuits may open the door to hybrid platforms that leverage the strengths of both axial and transverse photonic confinement for broader functional integration.

In conclusion, this work addresses limitations in tunable photonic delay lines, offering a simple, robust, and highly compact solution with relevance to optical beamforming, frequency comb stabilization, and neuromorphic photonic computing. The results confirm the feasibility of bent SNAP architectures as a flexible platform for dispersion-engineered slow-light photonic devices.

**Funding.** This research was supported by the Leverhulme Trust (grant RPG-2022-014) and the Engineering and Physical Sciences Research Council (EPSRC) under grants EP/W002868/1 and EP/X03772X/1.

**Disclosures.** The authors declare no conflicts of interest.

**Data availability.** Data underlying the results presented in this paper are not publicly available at this time but may be obtained from the authors upon reasonable request.